\documentclass[10pt, conference]{IEEEtran}

\usepackage{bm} 

\usepackage{amsmath}
\usepackage{graphicx}
\usepackage{url}
\usepackage{times}
\usepackage{booktabs}
\usepackage{tabularx}
\usepackage{hyperref}
\usepackage{scalerel}
\usepackage{tikz}
\usepackage{amssymb}
\usepackage{bbm}
\usetikzlibrary{svg.path}
\usepackage{siunitx}
\usepackage{tcolorbox}

\title{Data-Balanced Curriculum Learning for Audio Question Answering}

\author{\IEEEauthorblockN{Gijs Wijngaard, Elia Formisano, Michele Esposito, Michel Dumontier}
\IEEEauthorblockA{Maastricht University}}

\begin{document}

\nocite{IEEEexample:BSTcontrol}

\maketitle

\begin{abstract}
Audio question answering (AQA) requires models to understand acoustic content and perform complex reasoning. Current models struggle with dataset imbalances and unstable training dynamics. This work combines curriculum learning with statistical data balancing to address these challenges. The method labels question difficulty using language models, then trains progressively from easy to hard examples. Statistical filtering removes overrepresented audio categories, and guided decoding constrains outputs to valid multiple-choice formats. Experiments on the DCASE 2025 training set and five additional public datasets show that data curation improves accuracy by 11.7\% over baseline models, achieving 64.2\% on the DCASE 2025 benchmark.
\end{abstract}


\section{Introduction}
\label{sec:intro}

Audio Question Answering (AQA) represents a fundamental challenge at the intersection of acoustic understanding and natural language processing. Unlike simple sound classification, AQA requires models to comprehend complex acoustic scenes, identify temporal relationships between sounds, and generate coherent answers to diverse questions \cite{liMultiscaleAttentionAudio2023, beheraMultiLingualAudioQuestion2023, fayekTemporalReasoningAudio2020}. While earlier AQA models focused on binary tasks or limited label sets \cite{liMultiscaleAttentionAudio2023, sudarsanamAttentionBasedMethodsAudio2023}, later models contain answer generation capabilities in natural language beyond strict answer sets. Some tasks require text generation metrics to evaluate the output of the model, while other tasks require selecting the correct answer from provided options, which makes answers verifiable against actual answers. 

Audio-language modeling faces limited diversity and quality of training data \cite{wijngaardAudioLanguageDatasetsScenes2025}. Many models train on overlapping datasets, and this redundancy also appears in benchmarks in the form of data contamination, resulting in homogeneous dataset landscapes. This limits the generalization and robustness of audio-language models. Analysis reveals that audio datasets suffer from severe class imbalances. Certain sound categories show dramatic over-representation. This leads to models that perform well on common sounds but fail on rare acoustic events.

While traditional approaches to audio question answering have relied on supervised learning, recent advances have explored reinforcement learning techniques \cite{liReinforcementLearningOutperforms2025}. Reinforcement learning offers potential advantages for complex reasoning tasks but introduces training instabilities. This work investigates how curriculum learning and guided decoding can stabilize reinforcement learning for audio understanding.

This work tests the following hypotheses for improving audio question answering:

\begin{enumerate}
    \item \textbf{Curriculum-guided reinforcement learning}: The model trains on easy examples first to establish reliable reward signals. Difficult samples enter training progressively. This method stabilizes learning dynamics.
    \item \textbf{Statistical data balancing}: Statistical thresholds identify and remove overrepresented categories, thereby balancing the training datasets.
    \item \textbf{Guided decoding}: Regular expressions constrain generation to valid multiple-choice answers (A, B, C, D).
    \item \textbf{Hybrid training}: The model trains with Supervised Fine-Tuning (SFT) to provide stable initialization. This is followed by Group Relative Policy Optimization (GRPO) through reward-based learning.
\end{enumerate}

Experiments on six datasets validate the approach. Data quality determines performance more than algorithmic complexity. The method achieves 64.2\% accuracy on DCASE 2025 Task 5: Audio Question Answering.

\section{Background}
\label{sec:related}

\subsection{Audio-Language Models}
\label{ssec:audiolang}

Large-scale audio-language models transform audio understanding. Early approaches combine audio encoder models with language model decoders \cite{meiAudioCaptioningTransformer2021, koizumiTransformerbasedAudioCaptioning2020}. Here, the language model's performance is dependent on the quality of the features it receives from the audio encoder. Qwen2-Audio \cite{chuQwen2AudioTechnicalReport2024b} and other recent models address this through unified architectures. The model processes a combination of speech, audio, and environmental sounds with a Whisper-large-v3 encoder and a Qwen-7B language model, and is used in this work as a foundational model. 

Examples of training strategies in the audio-language domain include: temporal progression from 30-second to 5-minute contexts that enable 3B models to outperform larger architectures \cite{ghoshAudioFlamingo22025}. Perception-before-understanding improves comprehension \cite{gongListenThinkUnderstand2024}. Multi-phase thinking, which allows for planning, captioning, reasoning, and summarization \cite{xieAudioReasonerImprovingReasoning2025}.

\subsection{Training Strategies for Audio Understanding}
\label{sec:training_strategies}

Curriculum learning is the process of first showing a language model examples that can be classified as easy or simple, and gradually feeding the model more difficult examples \cite{bengioCurriculumLearning2009, elmanLearningDevelopmentNeural1993}. It shows promise in various machine learning domains \cite{sovianyCurriculumLearningSurvey2022a}. For audio-language models specifically, curriculum learning has enabled performance improvements through various progressive strategies: temporal progression from 30-second to 5-minute audio contexts allows 3B parameter models to outperform larger architectures \cite{ghoshAudioFlamingo22025}, and learning the model first to perceive and then understand facilitates better general audio comprehension \cite{gongListenThinkUnderstand2024}. The SARI framework \cite{wenSARIStructuredAudio2025} shows that curriculum-guided GRPO with structured reasoning can improve performance on complex audio reasoning tasks. Our method differentiates from the SARI framework by uniquely combining curriculum learning with statistical data balancing techniques to address dataset imbalances, and using guided decoding to ensure the model generates valid multiple-choice answers.

Reinforcement learning for language models gains attention through RLHF (Reinforcement Learning from Human Feedback) \cite{zieglerFineTuningLanguageModels2020}. Group Relative Policy Optimization (GRPO) \cite{shaoDeepSeekMathPushingLimits2024} extends these ideas by eliminating the separate value model, which reduces memory requirements while maintaining training stability. The method optimizes for complex objectives defined as reward functions that are difficult to capture through supervised learning alone.

Recent work extends GRPO to audio language models with promising results. Training with reinforcement learning shows significant improvements with fine-tuning Qwen2-Audio-7B-Instruct compared to supervised fine-tuning \cite{liReinforcementLearningOutperforms2025}. Further applications with Qwen2.5-Omni achieved state-of-the-art performance of 71.3\% accuracy, suggesting gains primarily stem from enhanced text reasoning capabilities rather than improved audio processing \cite{rouditchenkoOmniR1YouReally2025}. Curriculum learning integrated with the GRPO framework enables training models progressively from easy to difficult examples, where structured four-part reasoning (planning, caption, reasoning, summary) combined with curriculum scheduling achieves superior performance at 67.08\% accuracy \cite{wenSARIStructuredAudio2025}. A fundamentally different approach of decomposing audio into interpretable semantic components enhances model understanding of the audio beyond simple audio-to-text mapping \cite{wijngaardAudSemThinkerEnhancingAudioLanguage2025}.

Dataset-level curation strategies for audio-language models have recently gained attention as researchers recognize the importance of addressing class imbalances inherent in large-scale audio datasets. AudioSet, with its hierarchical ontology of 527 classes, exhibits significant imbalances where classes such as Music and Speech are dramatically overrepresented \cite{gemmekeAudioSetOntology2017}. In addition, LLMs play an important role in quality assessment by evaluating LLM-generated captions against human-written standards. Captions rated below a threshold are filtered out to ensure that only high-quality data is retained \cite{kreukAudioGenTextuallyGuided2023}. This work identifies and removes the most prevalent categories based on their standard deviation from the mean distribution. This ensures that our reinforcement learning phase focuses on learning from diverse examples to improve generalization to rare acoustic events.

\subsection{Audio Question Answering Datasets}
\label{ssec:datasets}

Challenge rules from the DCASE Challenge Task 5 restrict dataset selection to 20 approved sources. Many datasets target other tasks such as audio captioning and audio generation rather than question answering. This work selects only datasets with multiple-choice formats or relevant question-answer pairs. Five extra datasets are used to train the models, in addition to the training set of the challenge:

\begin{itemize}
    \item \textbf{AVQA} \cite{yangAVQADatasetAudiovisual2022}: Contains 56,369 question-answer pairs from VGGSound (originally 57,335 before YouTube deletions). Questions span eight semantic types. The dataset includes audio and visual information. AVQA requires models to discover causal correlations between modalities. 
    \item \textbf{ClothoAQA} \cite{lippingClothoAQACrowdsourcedDataset2022}: Contains 35,838 questions from 1,991 environmental audio samples (15-30 seconds). Each audio has 6 questions with 3 annotator responses per question. Four questions require binary answers; two require single words. Different annotators create questions and answers. 
    \item \textbf{CompA-Order} \cite{ghoshCompAAddressingGap2024}: Contains 900 examples for compositional reasoning. The dataset tests temporal understanding between sounds.
    \item \textbf{TACOS} \cite{primusTACOSTemporallyalignedAudio2025}: Contains 12k audio recordings with 61,137 temporally-aligned captions. The dataset is designed to train temporal reasoning.
    \item \textbf{AudSem} \cite{wijngaardAudSemThinkerEnhancingAudioLanguage2025}: Synthesized from YouTube closed captions. Contains 210,288 multiple-choice questions across diverse acoustic scenes and events.
\end{itemize}

\textbf{Evaluation Datasets.} The DCASE 2025 Challenge provides three specialized evaluation datasets. \textbf{Part 1 (Bioacoustics QA)} contains 0.7K training and 0.2K development examples from 31 marine mammal species from the Watkins Marine Mammal Sound Database \cite{woodsholeoceanographicinstitutionWatkinsMarineMammal}. Audio duration spans 0.4 to 625 seconds. \textbf{Part 2 (Temporal Soundscapes QA)} includes 1K training and 0.6K development examples covering 26 sound classes, which test temporal reasoning about sound sequences, timestamps, and durations. Part 2 sources include NIGENS general sound events database \cite{trowitzschNIGENSGeneralSound2019}, L3DAS23 Challenge \cite{marinoniL3DAS232023IEEE}, and TAU Spatial Sound Events 2019 \cite{adavanneTAUSpatialSound2019}. \textbf{Part 3 (Complex QA)} contains 6.4K training and 1.6K development examples from AudioSet \cite{gemmekeAudioSetOntology2017} and Mira datasets \cite{juMiradataLargescaleVideo2024}. Each 10-second audio clip (16 kHz) requires multi-faceted reasoning across temporal, acoustic, and contextual dimensions.

\section{Method}
\label{sec:method}

The method combines four components: (1) Curriculum-based filtering controls learning progression. (2) Statistical balancing ensures dataset diversity. (3) Hybrid training combines supervised fine-tuning with reinforcement learning. (4) Guided decoding for structuring output generation during GRPO training.

\subsection{Curriculum-Based Data Filtering}
\label{ssec:curriculum}

Traditional curation removes difficult examples as noise. These samples provide valuable signals when introduced properly. Curriculum learning trains models on easy examples before difficult ones \cite{bengioCurriculumLearning2009, elmanLearningDevelopmentNeural1993}. To create a subset of easy examples for initial training, we use a small LLM, Microsoft's Phi-4-mini-instruct \cite{abdinPhi4TechnicalReport2024}, to score question difficulty (see Box \ref{box:prompt} for prompt):

\begin{tcolorbox}[colback=white, colframe=black, boxrule=0.25pt, title={Prompt to Phi-4-mini-instruct for difficulty assessment}, label={box:prompt}]
\texttt{Task: Rate the difficulty of this audio-based question on a scale from 0.0 (very easy) to 1.0 (very difficult).\\
\\
Consider factors like:\\
- Complexity of the question\\
- Required knowledge/understanding\\
- Ambiguity or clarity of the question\\
- Number of concepts involved\\
\\
Question: \{question\}\\
Choices: \{choices\}\\
Correct Answer: \{answer\}}
\end{tcolorbox}

After labeling, the samples are filtered by difficulty. Easy samples (difficulty $<$ 0.3) establish reliable GRPO reward signals, which prevent instability from ambiguous examples. Later stages incorporate all samples or focus on hard examples (difficulty $>$ 0.7). Best results come from training on easy samples, then the full dataset.

\subsection{Statistical Category Balancing}
\label{ssec:balancing}

Audio datasets exhibit severe category imbalances. Dataset curation addresses class imbalances in audio datasets to promote diversity across the training data. AudioSet contains 527 classes with severe imbalances: Music and Speech dominate \cite{gemmekeAudioSetOntology2017}. A language model assesses caption quality against a description of human standards, where low-scoring captions are filtered \cite{kreukAudioGenTextuallyGuided2023}. Phi-4-mini-instruct \cite{abdinPhi4TechnicalReport2024} categorizes audio questions into 24 categories.

\begin{figure*}[ht!]
    \centering
    \includegraphics[width=1.0\textwidth]{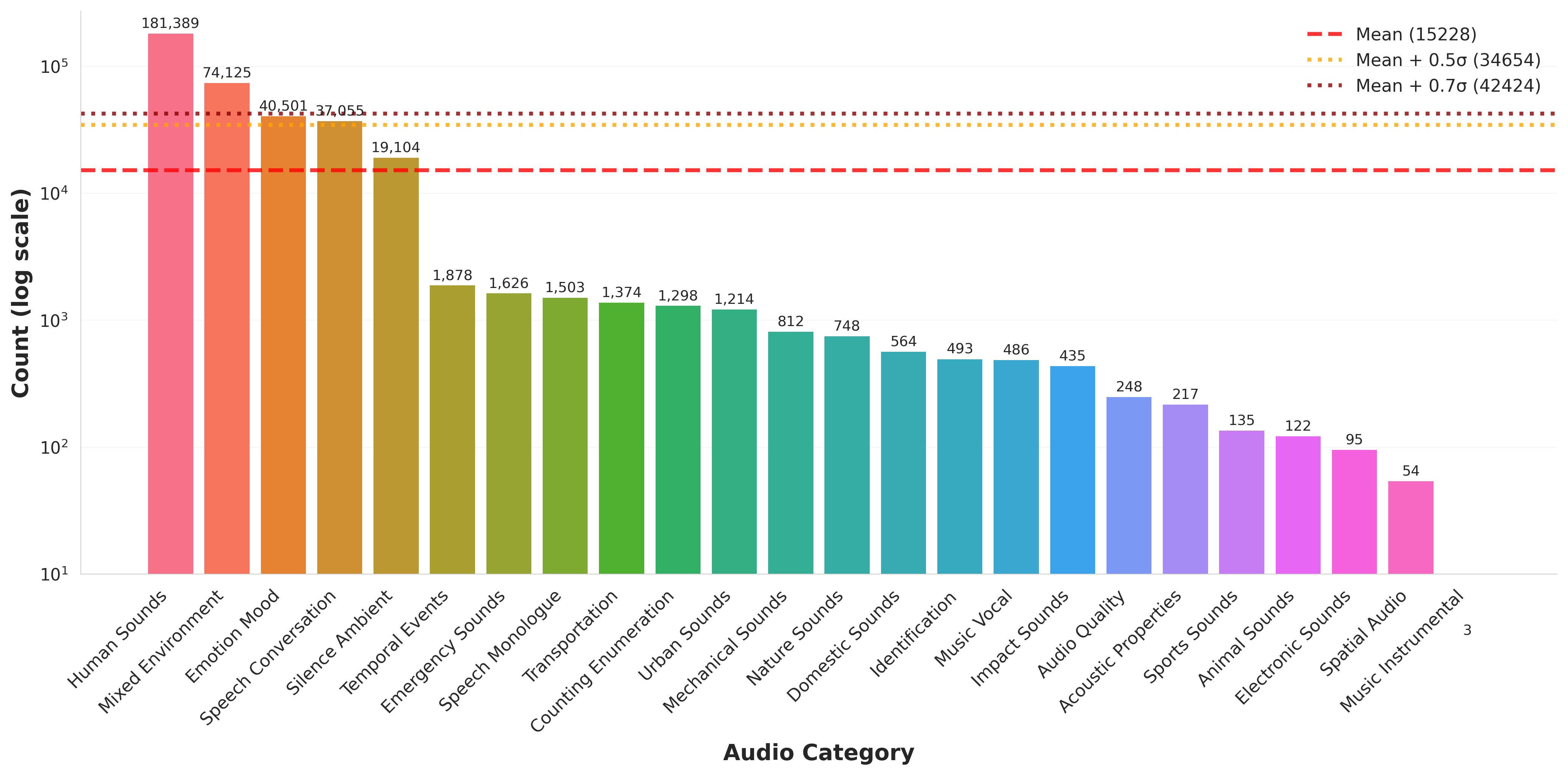}
    \caption{Distribution of audio categories in the training sets. Dotted lines represent the threshold during diversity filtering, with the cut-offs being $0.5 \sigma$ and $0.7 \sigma$ above the mean.
    }
    \label{fig:audio_categories}
\end{figure*}

In the filtering process, statistical thresholding balances category distributions. The method computes mean count $\mu$ across categories, and categories with count $c_i > \theta \cdot \mu$ get filtered. Optimal performance occurs at $\theta = 0.7$. This threshold removes extreme over-representation of human sounds and mixed environment sounds while maintaining sufficient examples. The approach prevents overfitting to dominant categories. In Figure \ref{fig:audio_categories}, the audio categories are visualized with their respective count and the cut-off threshold. The figure shows that human sounds and mixed environment sounds are over-represented in the training sets.

\subsection{Hybrid SFT-GRPO Training Pipeline}
\label{ssec:hybrid_training}

The training pipeline combines supervised fine-tuning and reinforcement learning. Group Relative Policy Optimization (GRPO) \cite{shaoDeepSeekMathPushingLimits2024} is a training technique introduced recently for reward-based optimization in finetuning models such as DeepSeek R1 \cite{shaoDeepSeekMathPushingLimits2024} and Qwen2.5 \cite{yangQwen25TechnicalReport2025}. Supervised Fine-Tuning (SFT) works well as a warm-up stage before GRPO, to adapt the model to the proposed formatting and style. Qwen2-Audio serves as the base model.

\textbf{Stage 1: Supervised Fine-Tuning (SFT)}. The method fine-tunes Qwen2-Audio-7B-Instruct using cross-entropy loss. Low-Rank Adaptation (LoRA) \cite{huLoRALowrankAdaptation2022} targets query and value projections with rank $r=8$, scaling factor $\alpha=16$, and dropout 0.05. Training uses AdamW optimizer \cite{loshchilovDecoupledWeightDecay2019a}, learning rate $2 \times 10^{-5}$, cosine annealing, and gradient clipping at 0.5. Training runs 1-3 epochs. 

\textbf{Stage 2: Group Relative Policy Optimization (GRPO)}. GRPO optimizes performance through reinforcement learning with verifiable rewards (RLVR) after SFT convergence. The reward function combines accuracy and format validation:

\[R(y, \hat{y}) = R_{acc}(y, \hat{y}) + R_{format}(\hat{y})\]

$R_{acc}$ awards 0.5 for full match, 0.25 for matching the answer letter (e.g. A, B, C, or D), and 0.25 for remaining content. $R_{format}$ awards 0.5 for correct format. GRPO uses $\beta = 0.01$ with a warmup of 50 steps, 4 generations per example, for 1-2 epochs, depending on the configuration. Dropout is disabled and the learning rate is $1 \times 10^{-6}$ with AdamW with cosine annealing. 

\textbf{Stage 3: Multi-Stage Pipeline}. Effective training chains three stages: (1) SFT on all datasets with LoRA. (2) GRPO on easy samples with diversity balancing. (3) GRPO on the full dataset with diversity balancing. Each stage builds on previous checkpoints. The pipeline combines SFT initialization with progressive GRPO improvements.

\subsection{Guided Decoding with Constrained Outputs}
\label{ssec:guided_decoding}

Models must produce valid multiple-choice answers. Guided decoding constrains a language model's output to ensure it conforms to a specific structure, using a regular expression \cite{willardEfficientGuidedGeneration2023a}. This is achieved by compiling the desired structure into a finite state machine, which can be visualized as a graph where each node represents a valid state in the grammar. At each step of text generation, the model can only transition to a new state (i.e., generate a next token) if there is a valid edge from its current position in the graph. This process works by creating a ``logit mask'' that effectively sets the probability of all invalid next tokens to zero, forcing the model to choose only from the tokens that keep the output syntactically correct according to the graph's rules. Constrained decoding ensures correct output format during generation.

Regular expressions constrain each decoding step. Valid paths correspond to answers A, B, C, or D. The regular expression \texttt{\^{}<think>.*?</think>\textbackslash s*<answer>(A|B|C|D).* </answer>\$} restricts the model's generation to a valid answer choice, while also allowing for reasoning. This approach of guided decoding eliminates post-processing since mathematical constraints force selection from four answer choices.

\subsection{Experimental Setup}
\label{ssec:expsetup}

\textbf{Datasets}: AVQA (56,369 QA pairs), ClothoAQA (35,838 questions), CompA-Order (900 examples), TACOS (61,137 captions), and AudSem (210,288 questions). \textbf{Baselines}: Qwen2-Audio, AudioFlamingo 2, and Gemini-2.0-Flash. \textbf{Metrics}: Top-1 accuracy with exact match evaluation. Results report overall accuracy and part-wise performance (Part 1: bioacoustics, Part 2: temporal/counting, Part 3: complex reasoning). \textbf{Evaluation:} Evaluation is done on the DCASE 2025 Task 5 evaluation set. Experiments compare against baseline models and conduct ablation studies. In total, 22 experimental runs are conducted, comparing different training strategies.

\section{Results}
\label{sec:results}
\begin{table}[ht!]
    \centering
    \caption{Accuracy on the DCASE 2025 evaluation set. Best results per column in bold.}
    \label{tab:main_results}
    \begin{tabular}{lcccc}
    \toprule
    \textbf{Model} & \textbf{Part 1} & \textbf{Part 2} & \textbf{Part 3} & \textbf{Total} \\
    \midrule
    \multicolumn{5}{l}{\textit{Baseline Models}} \\
    Qwen2-Audio-7B & 30.0\% & 39.2\% & 49.6\% & 45.0\% \\
    AudioFlamingo2 & 53.9\% & 31.7\% & 49.5\% & 45.7\% \\
    Gemini-2.0-Flash & 42.0\% & 46.3\% & 56.6\% & 52.5\% \\
    \midrule
    \multicolumn{5}{l}{\textit{Our Methods}} \\
    SFT only & 62.1\% & \textbf{42.2\%} & 72.5\% & 64.1\% \\
    GRPO + Curriculum & 67.0\% & 38.3\% & \textbf{72.8\%} & 63.7\% \\
    GRPO + Div. ($\theta=0.7$) & 65.6\% & 41.4\% & 72.6\% & \textbf{64.2\%} \\
    Ensemble (11 models) & \textbf{75.0\%} & \textbf{42.2\%} & 68.3\% & 62.5\% \\
    \bottomrule
    \end{tabular}
\end{table}

Table \ref{tab:main_results} presents the evaluation results on the DCASE 2025 test set submitted to the DCASE Challenge. The four training configurations are:

\begin{itemize}
    \item \textbf{SFT only}: A baseline model trained with supervised fine-tuning on all datasets for one epoch using LoRA optimization with rank 8 and learning rate $2 \times 10^{-5}$.
    \item \textbf{GRPO + Curriculum}: A three-stage pipeline that first performs SFT, then GRPO training on easy samples (difficulty $<$ 0.3) with diversity balancing at threshold 0.5, followed by GRPO on the full DCASE2025 training dataset for 5 epochs with guided decoding constraints. 
    \item \textbf{GRPO + Div. ($\theta=0.7$)}: Combines 3-epoch SFT training with aggressive diversity filtering that removes overrepresented categories when their count exceeds 0.7 times the mean category count, followed by two GRPO stages with the same diversity threshold.
    \item \textbf{Ensemble}: A majority voting ensemble combining all three individual models plus eight additional variants trained with different curriculum strategies and diversity thresholds. 
\end{itemize}  

Part 2 (temporal and counting tasks) remains the most challenging with accuracies between 38-42\%, which indicates fundamental difficulties in temporal reasoning within audio-language models. Part 1 (bioacoustics) and Part 3 (complex reasoning) show stronger performance, achieving 75.0\% and 72.8\% in the best scoring models respectively.

\subsection{Ablation of Model Components}
\label{ssec:ablation}
\begin{table}[ht!]
\centering
\caption{Ablation study results showing the impact of different components on DCASE 2025 test set accuracy. All ablations are built upon the SFT+GRPO training pipeline, which serves as our internal baseline (63.1\%).}
\label{tab:ablation}
\begin{tabular}{lc}
\toprule
\textbf{Configuration} & \textbf{Accuracy} \\
\midrule
\multicolumn{2}{l}{\textit{Training Paradigm}} \\
Pre-trained Qwen2-Audio (no fine-tuning) & 45.0\% \\
SFT only (single dataset) & 64.0\% \\
SFT only (all datasets) & 63.9\% \\
SFT + GRPO & 63.1\% \\
\midrule
\multicolumn{2}{l}{\textit{Diversity Balancing}} \\
No diversity filtering & 63.1\% \\
Diversity $\theta=0.3$ & 63.2\% \\
Diversity $\theta=0.7$ & 64.2\% \\
\midrule
\multicolumn{2}{l}{\textit{Curriculum Learning}} \\
No curriculum & 63.1\% \\
Easy samples (difficulty $<$ 0.2) & 64.2\% \\
Hard samples (difficulty $>$ 0.7) & 63.1\% \\
\midrule
\multicolumn{2}{l}{\textit{Additional Techniques}} \\
LoRA in GRPO & 62.3\% \\
Guided decoding & 63.7\% \\
\bottomrule
\end{tabular}
\end{table}

Table \ref{tab:ablation} summarizes an ablation study isolating the impact of different training strategies and data curation techniques.

Supervised fine-tuning with LoRA optimization forms the foundation of this approach. The pre-trained Qwen2-Audio model achieves 45.0\% accuracy on DCASE 2025. SFT with LoRA on a single dataset increases accuracy to 64.0\%. Expanding to all six datasets yields 63.9\% accuracy, which indicates dataset diversity alone provides minimal improvement without proper curation.

The multi-stage SFT+GRPO approach maintains 63.1\% accuracy, showing modest benefits compared to SFT alone. Notably, the ablation results reveal that GRPO does not outperform SFT when used alone, suggesting that reinforcement learning's effectiveness is conditional on data curation techniques that provide cleaner, more stable reward signals. Curriculum learning on easier examples (difficulty $<$ 0.2) achieves 64.2\%, while focusing on hard examples (difficulty $>$ 0.7) reduces performance to 63.1\%.

Diversity balancing produces the most significant improvements. Conservative filtering ($\theta=0.3$) yields 63.2\% accuracy, while aggressive filtering ($\theta=0.7$) achieves the best single-model result of 64.2\%. This confirms that removing overrepresented audio categories contributes more than simply adding data. Guided decoding experiments show modest improvement, as the model naturally learns to generate well-formatted outputs during training.

\subsection{Reasoning Phase}
Experiments incorporating a reasoning phase, where models generate intermediate reasoning steps before producing final answers, yielded significantly lower results (46.2-51.6\% accuracy). This limitation stems from dataset constraints: only AudSem among the six datasets includes thinking annotations. While GRPO theoretically enables thinking across all datasets by evaluating only final answers, poor SFT initialization from limited thinking examples prevents effective learning. Thinking-based approaches require comprehensive annotated reasoning across diverse datasets to compete with direct answer generation methods.

\section{Conclusion}
\label{sec:conclusion}

The method makes three contributions: (1) Statistical diversity balancing with threshold 0.7 addresses dataset imbalances. (2) Curriculum learning improves specific question types without significant overall gains. (3) Supervised fine-tuning with LoRA optimization delivers better performance than reinforcement learning alone, with SFT outperforming the SFT+GRPO approach unless combined with proper data curation techniques.

GRPO limitations highlight fundamental differences between audio question answering and text generation tasks. Future work requires alternative reward formulations or denser feedback mechanisms. The success of diversity filtering motivates more sophisticated curation approaches using learned representations instead of predefined categories.

Dataset quality matters more than algorithmic complexity in audio question answering. Effective data curation proved more impactful than the choice between advanced training algorithms for audio question answering. Diversity filtering provides the largest performance gains, achieving 64.2\% accuracy on DCASE 2025 Task 5. Diversity filtering applies to any audio-language dataset with minimal computational overhead. Balanced acoustic representation ensures more robust and equitable real-world performance across sound categories.

\section*{Acknowledgment}
This work was supported by the Dutch Research Council (NWO 406.20.GO.030 to Prof. Elia Formisano), the Dutch national e-infrastructure with the support of the SURF Cooperative using grant no. EINF-12157, Data Science Research Infrastructure (DSRI; Maastricht University) and the Dutch Province of Limburg.

\clearpage

\bibliographystyle{IEEEtran}
\bibliography{IEEEtranBSTCTL,refs}

\end{document}